\documentstyle[preprint,prb,aps]{revtex}   
 
\begin{document}   

\title{Electron scattering on disordered double barrier GaAs-Al$_x$Ga$_{1-x}$As
heterostructures}  

\author{I.\ G\'{o}mez,  E.\ Diez and F.\ Dom\'{\i}nguez-Adame}

\address{GISC, Departamento de F\'{\i}sica de Materiales, Universidad
Complutense, E-20840 Madrid, Spain}

\author{P.\ Orellana}

\address{Departamento de F\'{\i}sica, Universidad Cat\'olica del Norte, 
Casilla 1280, Antofagasta, Chile}

\date{\today}   
   
\maketitle

\begin{abstract} 
 
We present a novel model to calculate vertical transport properties such as
conductance and current in unintentionally disordered double barrier
GaAs-Al$_x$Ga$_{1-x}$As heterostructures. The source of disorder comes from
interface roughness at the heterojunctions (lateral disorder) as well as spatial
inhomogeneities of the Al mole fraction in the barriers (compositional
disorder). Both lateral and compositional disorder break translational symmetry
along the lateral direction and therefore electrons can be scattered off the
growth direction. The model correctly describe channel mixing due to these
elastic scattering events. In particular, for realistic degree of disorder, we
have found that the effects of compositional disorder on transport properties
are negligible as compared to the effects due to lateral disorder. 

\end{abstract}   
   

\narrowtext

\section{Introduction}

Resonant tunneling (RT) through double-barrier structures (DBS) make these
systems very promising candidates for a new generation of ultra-high speed
electronic devices. For instance, a GaAs-Ga$_{1-x}$Al$_x$As DBS operating at
THz frequencies has been reported in the literature.\cite{Sollner84} The basic
reason for RT to arise in DBS is a quantum phenomenon whose fundamental
characteristics are by now well understood: There exists a dramatic increase of
the electron transmission whenever the energy of the incident electron is close
to one of the unoccupied quasi-bound-states inside the well.\cite{Ricco84} In
practice, a bias voltage is applied to shift the energy of this
quasi-bound-state of nonzero width so that its center matches the Fermi level. 
Consequently, the $j-V$ characteristics present negative differential
resistance when the quasi-bound-state lies well below the Fermi level.

In actual samples, however, the situation is much more complex than this simple
picture. Scattering by phonons, electrons or defects reduces the required
quantum coherence and, in fact, deviations from the above simple description
are observed. These scattering mechanisms explain the occurrence of side-band
resonant peaks due to the interaction with phonons\cite{Wingreen88,Cai89} and
photons,\cite{Chitta94,Inarrea94} hysteresis in the $j-V$ characteristics due
to many body effects,\cite{Levy93,Orellana99,Orellana00} and the  decrease of
electron mobility due to rough surfaces even in good-quality
heterostructures.\cite{Penner98} While one-dimensional models successfully
describe scattering by electrons and phonons, they obviously cannot account for
all interface roughness effects. So far, there are analytical results
concerning the propagation of wave packets in a randomly layered medium when
the potential is a random function of only one coordinate,\cite{Freilikher91}
but for a small number of layers, as in DBS, in-plane disorder becomes
important and one expects such approaches to fail. Realistic models of in-plane
disorder, including power-like or exponential spatial correlations observed by
transmission electron microscopy, usually lead to intractable analytical
models; hence the importance of numerically solvable models to bridge this
gap. An important contribution was already provided by Henrickson {\em et al.},
who applied the tight-binding Green function method to interface roughness 
in DBS.\cite{Henrickson92} In this paper transmission was studied for a DBS with
a rather simple model of disorder (periodic roughness with random relative 
phases at the interfaces). A level splitting was found for narrow well DBS and
a dependence on the roughness lateral size was observed.

In this work we present a simple two-dimensional model that allowed us to carry
out an extensive study of electron scattering by interface roughness and how it
manifests trough vertical transport properties of the DBS. In particular, we
aim to elucidate the relationship between macroscopic properties of the DBS
(e.g.\@ conductance) with microscopic parameters of the model (e.g.\@
correlation length of the surface disorder). The paper is organized as follows: 
The body of the paper is Sec.~\ref{sec2}, where we present our model based, on
the effective-mass approximation, describing a disordered sample (e.g.\@ an
imperfect DBS) connected to two perfectly ohmic leads. The Ben Daniel-Duke
equation is discretized, boundary conditions are discussed and scattering
solutions are found by means of the transfer-matrix method for {\em any\/}
arbitrary heterostructure made of wide gap semiconductors. Afterwards,
expressions for two-terminal conductance and current for unintentionally
disordered DBS are provided. The model is
worked out in a two-dimensional space for computational limitations, although
it will be clear that generalization to three dimensions is rather
straightforward. Section~\ref{sec3} describes the various models  of disorder
we have used to mimic structural data obtained by transmission electron
microscopy and X-ray scattering. In Sec.~\ref{sec4} we present the numerical
results and, in particular, the relationship between macroscopic properties of
the DBS with microscopic parameters of the model. Finally, in Sec.~\ref{sec5}
we discuss our results and the main conclusions of the work.

\section{Model} \label{sec2}

\subsection{Discrete equation}

We consider the single-electron two-dimensional Schr\"{o}dinger equation
within  the framework of the one-band effective-mass approximation. Close to
the $\Gamma$ valley, this approach leads to the Ben Daniel-Duke equation for
the envelope function
\begin{equation} 
\label{hamiltonian}
\left[-\frac{\hbar ^{2}}{2m^{*}}\left( \frac{\partial^{2}}{\partial y^{2}}
+\frac{\partial^{2}}{\partial z^{2}}\right)+U(y,z) \right]\psi(y,z)=E\psi(y,z).
\end{equation}
where $z$ denotes the coordinate along the growth direction (see
Fig.~\ref{fig1}). Notice that we have taken a constant effective mass $m^{*}$
at the $\Gamma$ valley although this is not a serious limitation as the model
can be easily generalized to include a position-dependent effective mass. We
then consider a mesh with lattice spacings $ a_y$ and  $a_z$ in the $y$ and $z$
directions, respectively. Defining $t_y \equiv  -\hbar^{2}/(2m^{*}a_y^2)$ and
$t_z\equiv -\hbar^{2}/(2m^{*}a_z^2)$, we arrive at the following discretized
version of Eq.~(\ref{hamiltonian})
\begin{equation} 
\label{discrete}
t_z(\psi_{n+1,m}+\psi_{n-1,m})+t_y(\psi_{n,m+1}+\psi_{n,m-1}) 
+(U_{n,m}-2t_z-2t_y)\psi_{n,m} = E\psi_{n,m}.
\end{equation}
The potential term $U_{n,m}$ in Eq.~(\ref{discrete}) is given by the
conduction-band edge energy at the point $(na_y,ma_z)$ which, in turn, depends
on the Al mole fraction in the vicinity of that position. Consequently, both
kinds of disorder (lateral and compositional) are taken into account through
this  diagonal term (diagonal disorder). 

\subsection{Transfer matrix formalism}

In order to solve the tight binding-like equation~(\ref{discrete}) we use the
transfer matrix method based on the solutions calculated for each slide of the
system along the $z$ direction. To this end, we define 
\begin{mathletters}
\begin{equation} 
\label{vectors}
\phi_n \equiv \left(
\begin{array}{c}
\psi_{n,1} \\
\psi_{n,2} \\
\vdots \\
\psi_{n,M}
\end{array}
\right),
\quad n=0,1,\ldots,N+1,
\end{equation}
and
\begin{equation}
\cal M \em _n = \left(
\begin{array}{ccccc}
U_{n,1}-2t_z-2t_y & t_y & 0 & \cdots & t_y \\
t_y & U_{n,2}-2t_z-2t_y & t_y & \cdots & 0   \\
0 & t_y & U_{n,3}-2t_z-2t_y & \cdots & 0   \\
\vdots & \vdots & \vdots & \ddots & \vdots \\
t_y & 0 & 0  & \cdots & U_{n,M}-2t_z-2t_y 
\end{array}
\right).
\end{equation}
\end{mathletters}
Here $M$ and $N+1$ are the number of mesh divisions in the $y$ and $z$
directions, respectively. Note that periodic boundary conditions have been
taken into account on each slide. With these definitions, Eq.~(\ref{discrete})
can be cast in a more compact form
\begin{equation} 
\label{promotion}
\left(
\begin{array}{c}
\phi_{n-1} \\
\phi_{n}
\end{array} 
\right) =
\left(
\begin{array}{cc}
t_{z}^{-1}(E \cal I \em - \cal M \em _n) & - \cal I \em \\
\cal I \em & \cal O \em
\end{array} 
\right)
\,
\left(
\begin{array}{c}
\phi_{n} \\
\phi_{n+1}
\end{array}
\right)
\end{equation}
where $\cal I$ and $\cal O$ are the $M\times M$ identity and null matrices
respectively. This expression allows us to relate by iteration the envelope
function  amplitudes at non-consecutive slides. In particular, we can obtain
the expression of the envelope function amplitudes in the left contact as a
function of the amplitudes in the right one
\begin{mathletters}
\begin{equation} 
\label{transfer}
\left(
\begin{array}{c}
\phi_{0} \\
\phi_{1}
\end{array} 
\right) =
{\cal T}^{(N)}
\,
\left(
\begin{array}{c}
\phi_{N} \\
\phi_{N+1}
\end{array}
\right),
\end{equation}
where
\begin{equation} 
{\cal T}^{(N)} \equiv \prod_{n=1}^{N}
\left(
\begin{array}{cc}
t_{z}^{-1}(E \cal I \em - \cal M \em _n) & - \cal I \em \\
\cal I \em & \cal O \em
\end{array} 
\right).
\end{equation}
\end{mathletters}
is the transfer matrix for the heterostructure.

\subsection{Scattering solutions}

We are interested in the scattering of electrons incident from the left
contact. The envelope functions within the contacts will be determined by the
boundary conditions. These boundary conditions are open in the $z$ direction,
and periodic on each slide, that is in the $y$ direction. The former imply
plane wave solutions in the $z$ axis, and the latter yield an energy
discretization on $y$. As a consequence, this discretization results in a
number of transverse channels equal to the number of points in the transverse
mesh direction. Considering by the moment 
$U_{n,m}=0$ at the contacts and setting
$\psi_{n,1}=\psi_{n,M+1}$, a particular solution of Eq.~(\ref{discrete}) is
given by
\begin{equation} 
\label{particular}
\psi_{n,m}^l = \frac{1}{\sqrt{{\cal N}_l}}\exp\left(i\,\frac{2\pi l}{M}m\right) 
\exp\left(i k_l a_z n\right), \quad 
l=1,2,\ldots,M.
\end{equation}
Here ${\cal N}_l$ is a normalization constant that is needed for the different
{\em propagating\/} modes to carry the same current all of them. This
normalization constant is given by
\begin{equation} 
\label{normalization}
{\cal N}_l= \frac{1}{a^2_y}\,\sin^2 \left( \frac{2 \pi l}{M} \right)
+\frac{1}{a^2_z}\,\sin^2 \left( k_l a_z \right)
\end{equation}
Note that we consider a dependence on the index $l$ of the momentum in the
$z$ direction. This is a consequence of the implicit assumption that all the 
scattering processes in region~II are elastic, so energy is conserved. Thus, 
for a given energy 
\begin{equation} 
\label{kl}
k_l = \frac{1}{a_z} \cos^{-1} \left\{ \frac{1}{2t_z} \left[ E - 2t_y 
\left( \cos \frac{2\pi l}{M} -1 \right) \right] + 1 \right\}.
\end{equation}
Finally we can write a general solution of the scattering problem for an
electron of energy $E$ impinging on the scattering region on the $l$ transverse
mode. On the left contact the solution will be of the form
\begin{mathletters}
\begin{eqnarray} 
\label{solutionI}
\psi_{n,m}^{l}&=&\frac{1}{\sqrt{{\cal N}_l}}
\exp\left(i\,\frac{2\pi l}{M}m\right)
\exp\left(ik_l a_z n\right) \nonumber \\
&+& \sum_{j=1}^{M}\widehat{r}_{lj} \frac{1}{\sqrt{{
\cal N}_j}}\exp\left(i\,\frac{2\pi j}{M}m\right)
\exp\left(-ik_ja_z n\right), \quad (m,n) \in \mathrm{I}.
\end{eqnarray}
On the right contact we have
\begin{equation} \label{solutionIII}
\psi_{n,m}^{l}=
\sum_{j=1}^{M} \widehat{t}_{lj} \frac{1}{\sqrt{{\cal N}_j}}\exp\left(i\,
\frac{2\pi j}{M}m\right) \exp\left(i k_j a_z n\right),
\quad (m,n) \in \mathrm{III}.
\end{equation}
\end{mathletters}
The matrices $\widehat{r}$ and $\widehat{t}$ appearing in these two solutions
are the {\em reflection\/} and {\em transmission\/} matrices and they are
responsible for the channel {\em mixing\/} due to scattering events. Thus,
$\widehat{r}_{ij}$ represents the probability amplitude for an electron
impinging in channel $i$ to be reflected into the chanel $j$. Similarly
$\widehat{t}_{ij}$ represents the  probability amplitude for an electron
impinging in chanel $i$ to be transmited through the scattering region II into 
the channel $j$. Note that the solution  within region II is unknown. Actually, 
we are not interested in this solution since all the physics of the scattering
problem is contained in the {\em mixing\/} matrices $\widehat{t}$ and
$\widehat{r}$. In the following, our main goal will be to relate the elements
of $\widehat{t}$ and $\widehat{r}$ to those of ${\cal T}^{(N)}$ in
Eq.~(\ref{transfer}).

We start by re-writting formally Eq.~(\ref{discrete}) as ${\cal H} \psi = E
\psi$. Now we perform the following transformation upon the envelope functions 
\begin{equation} 
\label{transformation}
\widetilde{\psi} = \widehat{t}^{-1} \psi.
\end{equation}
It is easy to see that the Hamiltonian is invariant under such a
transformation, that is, $\widetilde{\cal H} \equiv \widehat{t}^{-1} {\cal H}
\widehat{t} = {\cal H}$. This is clear since $\widehat{t}$ and ${\cal H}$
formally commute as they act upon different vector spaces. The fact that ${\cal
H}$ is invariant implies that ${\cal T}^{(N)}$ is not affected by
transformation~(\ref{transformation}) so it remains unchanged.
In addition, equations~(\ref{solutionI}) and~(\ref{solutionIII}) are
transformed into
\begin{mathletters}
\begin{eqnarray} 
\label{new_solutionI}
\widetilde{\psi}_{n,m}^{l} & = & \sum_{j=1}^{M} \widehat{a}_{lj} 
\frac{1}{\sqrt{{\cal N}_j}} \exp\left(i\,\frac{2\pi j}{M}m\right)
\exp\left(i k_j a_z n\right)\nonumber \\ & + &
\sum_{j=1}^{M} \widehat{b}_{lj} \frac{1}{\sqrt{{\cal N}_j}} \exp\left(i\,
\frac{2\pi j}{M}m\right) \exp\left(-i k_j  a_z n\right),
\quad (m,n) \in \mathrm{I}
\end{eqnarray}
and
\begin{equation} 
\label{new_solutionIII}
\widetilde{\psi}_{n,m}^{l}=\frac{1}{\sqrt{{\cal N}_l}} \exp\left(i\,
\frac{2\pi l}{M}m\right) \exp\left(i k_l a_z n\right),
\quad (m,n) \in \mathrm{III},
\end{equation}
\end{mathletters}
respectively, where we have defined $\widehat{b}_{lj} \equiv \sum_{s=1}^{M}
\widehat{t}^{-1}_{ls} \widehat{r}_{sj}$ and $\widehat{a}_{lj} \equiv
\widehat{t}^{-1}_{lj}$. For each channel $l$ we can write the transform of
Eq.~(\ref{transfer}) as
\begin{equation} 
\label{chanel_l}
\left(
\begin{array}{c}
\widetilde{\phi}_{0}^l \\
\widetilde {\phi}_{1}^l
\end{array} 
\right) =
{\cal T}^{(N)}
\,
\left(
\begin{array}{c}
\widetilde{\phi}_{N}^l \\
\widetilde{\phi}_{N+1}^l
\end{array}
\right)\ .
\end{equation}
Now, introducing the definitions
\begin{eqnarray} 
\label{definitions}
{\cal A}^0 &\equiv& A^0_{jm}= \frac{1}{\sqrt{{\cal N}_j}} \exp\left(i\,
\frac{2\pi j}{M} m\right)\nonumber,\\
{\cal B}^0 &\equiv& {\cal A}^0, \nonumber \\
{\cal A}^1 &\equiv& A^1_{jm}= \frac{1}{\sqrt{{\cal N}_j}}\exp\left(i\,
\frac{2\pi j}{M} m \right) \exp\left(i k_j a_z\right), \nonumber  \\
{\cal B}^1 &\equiv& B^1_{jm}= \frac{1}{\sqrt{{\cal N}_j}}\exp\left(i\,
\frac{2\pi j}{M} m \right) \exp\left(- i k_j  a_z\right), \nonumber \\
\widehat{a}^l &=& \left( \begin{array}{c}
\widehat{a}_{l1} \\
\widehat{a}_{l2} \\
\vdots \\
\widehat{a}_{lM}
\end{array} \right), \quad
\widehat{b}^l = \left( \begin{array}{c}
\widehat{b}_{l1} \\
\widehat{b}_{l2} \\
\vdots \\
\widehat{b}_{lM}
\end{array} \right),
\end{eqnarray}
we arrive at the following transform of Eq.~(\ref{chanel_l}) in terms of the
elements of the {\em mixing\/} matrices
\begin{equation} 
\label{defin_I}
\left(
\begin{array}{c}
\widetilde{\phi}_{0}^l \\
\widetilde {\phi}_{1}^l
\end{array} 
\right) =
\left(
\begin{array}{cc}
{\cal A}^0 & {\cal B}^0 \\
{\cal A}^1 & {\cal B}^1 
\end{array}
\right) \,
\left(
\begin{array}{c}
\widehat{a}^l \\
\widehat{b}^l 
\end{array} 
\right) =
{\cal T}^{(N)}
\,
\left(
\begin{array}{c}
\widetilde{\phi}_{N-1}^l \\
\widetilde{\phi}_{N}^l
\end{array}
\right),
\end{equation}
and, finally, we have
\begin{equation} 
\label{defin_II}
\left(
\begin{array}{c}
\widehat{a}^l \\
\widehat{b}^l 
\end{array} 
\right) =
\left(
\begin{array}{cc}
{\cal A}^0 & {\cal B}^0 \\
{\cal A}^1 & {\cal B}^1 
\end{array}
\right)^{-1}
\, {\cal T}^{(N)} \,
\left(
\begin{array}{c}
\widetilde{\phi}_{N-1}^l \\
\widetilde{\phi}_{N}^l
\end{array}
\right),
\end{equation}
that give us the matrix elements $\widehat{a}_{lj}$ and $\widehat{b}_{lj}$
with  $j=1,2,\ldots,M$ for each channel $l$ in terms of the product of two
known matrices  and the transformed solutions~(\ref{new_solutionIII}) at the
right contact for the $l$ channel. It is worth noticing that all the scattering
information is contained in the transfer matrix ${\cal T}^{(N)}$. Thus, this
quantity turns out to be the fundamental object in the resolution  of the
scattering problem.

\subsection{Conductance and current}

Once $\widehat{a}$ and $\widehat{b}$ have been calculated, obtaining
$\widehat{t}$ and $\widehat{r}$ is an easy task. However, the matrices
calculated this way are the {\em response\/} matrices that contains some
non-physical information. This non-physical information comes from the fact
that we have been considering {\em all\/} the mathematical solutions of the
scattering problem, including those diverging at infinity. These solutions are
those for which momentum in Eq.~(\ref{kl}) is an imaginary number. In order to
avoid the unphysical solutions we will cancel out all the elements on
$\widehat{t}$ for which the incoming and/or the outgoing transverse quantum
number $l$ satisfy the following condition
\begin{equation} 
\label{physical_condition}
 E < 2 t_y \left( \cos \frac{2 \pi l}{M} -1 \right).
\end{equation}
Once the physical {\em mixing\/} matrices are known, particularly the
transmission matrix $\widehat{t}$, we can use them to calculate different
physical quantities  like the conductance or the electric current. From the
Landauer-B\"uttiker formalism,\cite{Landauer57} the zero temperature two-leads
multichannel conductance can be calculated using de Fisher-Lee
formula\cite{Fisher81} 
\begin{equation} 
\label{conductance}
G = \frac{2e^2}{h} \text{Tr} (\widehat{t}^{\dag}\widehat{t}).
\end{equation}
In order to calculate the electric current due to an applied field $F$ we have
to modify slightly our equations and further approximations are needed. First
of all, we consider perfect leads, so no voltage drop occurs within the
contacts. Besides, we will not consider the contribution to the current of
electrons impinging from the right contact. This is a good approximation
provided $eV > E_F$, where $V$ is the voltage drop along the region~I and $E_F$
is the electron Fermi level on the contacts. Finally, we will assume zero
temperature so no carrier statistics will be taken into account.

When an electric field applied along the growth direction  is considered, the
potential $U(y,z)$ in Eq.~(\ref{hamiltonian}) has to be replaced by
$U(y,z)+U_F(z)$, where $U_F$ is constant in the contacts ($U_F(z)=0$ in
region~I and $U_F(z)=-eV$ in region~III) and behaves linearly in region~II,
namely $U_F(z)=-eFz$. Here $F$ is the applied electric field and $V=FL$ is the
voltage drop across the scattering region whose length is $L$. The
electron momentum on the right contact now reads
\begin{equation} \label{ql}
q_l = \frac{1}{a_z} \cos^{-1} \left\{ \frac{1}{2t_z} \left[ E + eV - 2t_y 
\left( \cos \frac{2\pi l}{M} -1 \right) \right] + 1 \right\},
\quad (m,n) \in \mathrm{III}
\end{equation}
and solution in region~III changes to
\begin{equation} \label{new_solution}
\psi_{n,m}^{l}=\sum_{j=1}^{M} \widehat{t}_{lj} \frac{1}{\sqrt{{\cal N}_j}} 
\exp\left(i\,\frac{2\pi j}{M}m\right) \exp\left(i q_j a_z n\right).
\end{equation}

Bearing in mind all the approximations considered so far, a new transfer matrix
${\cal T}^{(N)}$ can be calculated for a each value of $F$. Inserting solution
(\ref{new_solution}) in Eq.~(\ref{defin_II}) we can obtain the {\em mixing\/}
matrices for a given applied field and then calculate using them the electric
current induced by that field. From the expression for the probability current
we can write the discretized version of the electronic current density
contribution of chanel $l$ along the applied field direction across the right
lead (calculated on a slide $n$ within the region~III) as
\begin{equation} \label{current}
j^l_z=i \frac{et_z a_z}{\hbar a_y} \frac{1}{M} \sum_{m=1}^M \left[
(\psi^l_{n,m})^* \psi^l_{n+1,m}-\psi^l_{n,m}(\psi^l_{n+1,m})^* \right].
\end{equation}
In order to calculate the total density current across the sample, we just need
to sum up over all the allowed transverse states below the Fermi level at the
left contact, that is
\begin{equation} \label{total_current}
j_z=i \frac{et_z a_z}{2 \pi \hbar a_y} \frac{1}{M} 
\sum_{m=1}^M \sum_{l=1}^M \int_0^{k_F^l} \left[
(\psi^l_{n,m})^* \psi^l_{n+1,m}- \psi^l_{n,m}(\psi^l_{n+1,m})^* \right] dk.
\end{equation}
Notice that the electron momentum in the $z$ direction, $k$, varies 
continously as the energy for an incident electron can take values from $0$ to 
$E_F$. Here, $k_F^l$ is the component of the electron momentum along the $z$ 
direction  for an incident electron with transverse quantum number $l$ and an 
incident energy $E_F$. Using the transmission matrix elements we finally arrive 
at
\begin{eqnarray} 
\label{final_current}
j_z&=&i \frac{et_z a_z}{2 \pi \hbar a_y} \frac{1}{M} 
\sum_{m=1}^M \sum_{l=1}^M \int_0^{k_F^l}
\sum_{j=1}^M \sum_{s=1}^M \widehat{t}^*_{lj} \widehat{t}_{ls} 
\frac{1}{\sqrt{{\cal N}_s{\cal N}_j}}
\exp\left(i\,\frac{2\pi(s-j)}{M}m\right) \nonumber \\
&\times& \exp\left[i(q_s-q_j) a_z n\right]
\left[\exp\left(iq_s a_z\right)- \exp\left(-iq_j a_z\right) \right] dk.
\end{eqnarray}

The results we have obtained so far provide an exact, although nonclosed,
analytical description of {\em any\/} two-dimensional heterostructure based on
wide-gap semiconductors, whenever the Ben Daniel-Duke holds valid. Let us
stress that the generalization to three-dimensional heterostructures is fairly
straightforward. With these results at hand, we can compute the transport
magnitudes we mentioned above. All expressions are very simple and suitable for
an efficient numerical treatment for any specific case. We will now evaluate
them for unintentionally disordered DBS to describe the relevant features of
the model.

\section{Models of disorder} \label{sec3}

Unintentional disorder appearing during growth in {\em actual\/}
heterostructures depends critically on the growth conditions. There exist
several techniques, like scanning tunneling
microscopy~\cite{Feenstra94,Salemik92,Gwo93} and X-ray
scattering,~\cite{Jergel98} which have been applied in recent years to
quantitatively determine structural properties of multilayers and
superlattices. Precise information about the nature and extend of defects at
interfaces is now available. Following M\"{a}der {\em et al.}, disorder in a 
heterostructure can be classified into two categories, namely lateral and
vertical.\cite{Mader95} Lateral disorder occurs whenever one semiconductor
protrudes into the other, forming chemically intermixed interfaces, steps and
islands. As a consequence, the interface is not flat and translational symmetry
in the plane perpendicular to the growth direction is broken. On the other
side, vertical disorder is observed whenever layer thicknesses fluctuate around
their nominal values. Compositional disorder due to different local Al 
concentration in each Ga$_{1-x}$Al$_{x}$As layer can be viewed as mixing of 
vertical and lateral disorder. 

\subsection{Lateral disorder}

In order to treat lateral disorder we have considered the formation of islands
on the interface between two consecutive layers having identical 
lateral sizes all of them and being consecutive one to each other. In our
model islands have heights that are randomly distributed (see
Fig.~\ref{fig2}). It is possible to express the rough profile of the interface 
between two consecutive layers defining the following height function
\begin{equation} \label{deviation}
h(y)= \eta \sum_n w_n \{ \Theta(y-n\zeta)+ \Theta [ (n+1)\zeta-y] -1\}.
\end{equation}
Here $h(y)$ represents the deviation from the flat surface at position $y$,
$\Theta$ is the Heavyside theta function, $\zeta$ is the island width, $w_n$ 
is a random variable associated to the $\em n\/$th island that controls the
fluctuation around the mean value, and $\eta$ is the largest deviation ---in
absolute value--- assuming that the $w_n$'s are uniformly distributed between
$-1$ and $1$. Hereafter $\eta$ will be referred to as degree of {\it lateral 
disorder}. In the following, two models will be introduced. First of all,
uncorrelated disorder will be considered, for which the random variables $w_n$
take values from $-1$ to $1$ uniformly, satisfying the following correlator
$\langle w_n w_m \rangle = \delta_{nm}/3$. However, experimental data and
theoretical models of molecular-beam-epitaxy growth
processes\cite{Barabasi95} indicate that the heigths of the surface at distant
points are strongly correlated. To mimic this situation, we will consider a
second model of lateral disorder where the $w_n$'s are normally distributed
with zero mean satisfying the
exponential correlator $\langle w_n w_m \rangle = \exp(-|n-m|/\xi)/3$, where
$\xi$ is the correlation length of the disorder. 

\subsection{Compositional disorder}

Compositional disorder in GaAs-Al$_x$Ga$_{1-x}$As heterostructures is due to
the lack of spatial uniformity of the Al mole fraction during the
heterostructure  growth process. We will simulate compositional disorder by
defining a spatial two-dimensional mesh over the Al$_x$Ga$_{1-x}$As barriers,
as shown in Fig.~\ref{fig3}. We will associate a value $x_{ij}$ for the Al
mole fraction to each region of the mesh. This value will fluctuate randomly
around the nominal value of the aluminum fraction on the barrier $\bar {x}$.
Thus we can write 
\begin{equation} \label{mole_fraction}
x_{ij}=\bar{x}+\Omega w_{ij}
\end{equation}
where $\Omega$ is the maximum fluctuation of the Al mole fraction, and $w_{ij}$
is a random variable related to the $(i,j)$ point of the mesh, taking values
uniformly between $-1$ and $1$. Note that the overall averaged  mole fraction
is equal to $\bar{x}$.

\section{Results} \label{sec4}
  
We have performed several numerical calculations in order to study the effect
of both lateral and compositional disorder over the transport properties of DBS
made of GaAs-Al$_x$Ga$_{1-x}$As heterostructures. 

\subsection{Uncorrelated lateral disorder}

We start by considering the effect of the interface roughness, characterized by
the degree of lateral disorder $\eta$ given in (\ref{deviation}), on the
conductance. Figure~\ref{fig4} shows the conductance calculated for different
values of $\eta$. Here $a_y=10\,$nm, $a_z=0.3\,$nm, $\zeta=20\,$nm, $M=50$, and
$N=38$.
The barrier widths are $2.1\,$nm for both the emitter and the collector, their 
heights are also the same, $0.3\,$eV, and the well width is $4.8\,$nm. The
curves correspond to an ensemble average of the conductance curves over $100$
different realizations of  the lateral disorder. Three different values of
$\eta$ were studied, namely $\eta=0$ (perfect DBS), $\eta=0.3\,$nm (largest
fluctuation of the order of one monolayer) and $\eta=0.6\,$nm (largest
fluctuation of the order of two monolayers). As a main result, it can be seen
in Fig.~\ref{fig4} that increasing the degree of lateral disorder, $\eta$,
results in a decrease of the conductance at the resonant energy.

Notice that the resonant conductance peak slightly widens due to the
fluctuations of its energy for each realization of the disorder. Besides, an
additional effect can be seen, that is, as the degree of lateral disorder
$\eta$ increases the conductance peak shifts to smaller energies. This effect
will be important in order to explain the $j-V$ characteristics later. Regarding
the effect of the size of the islands, $\zeta$, we have observed that it can be
neglected unless this size is of the order of the electron wavelength, that is,
for $\zeta \gg \lambda_e$ the conductance does not depend on $\zeta$.  As
expected, when $\zeta \sim \lambda_e$ the electron starts to {\em see\/} the
disorder and then the conductance increases as $\zeta$ decreases, as shown in
Fig.~\ref{fig5}. For energies about $0.1\,$eV this transition takes place at
sizes $\zeta \sim 10\,$nm~\cite{Henrickson92}. 

Current as a function of the applied bias is depicted in Fig.~\ref{fig6}. For
this calculation we have chosen $a_y=10\,$nm, $a_z=0.1\,$nm, $\zeta=10\,$nm,
$M=20$ and $N=77$. The barrier widths are $2.1\,$nm for both the emitter and the
collector and their heights are also the same, $0.25\,$eV. The well width is
$2.9\,$nm. Curves show in Fig.~\ref{fig6} correspond to an average over $50$
different realizations of the lateral disorder. The Fermi level was fixed at
$20\,$meV, the maximum applied bias was $0.5\,$eV, and $\eta=0.3$\ nm.
Surprisingly, current in disordered DBS is larger than that measured in
ordered DBS. This counterintuitive result can be explained recalling that
conductance peak shifted to lower energies, as mentioned above. The lowering of
the conductance peak can be understood assuming that surface roughness makes
the effective width of the quantum well larger than its nominal value. This
would imply a lower transmission resonance so the current peak shifts toward
lower bias. But an effective wider well implies a higher current too, so in
some statistical sense the current for the disordered DBS will be higher.

\subsection{Correlated lateral disorder}

Up to now we have been considering only lateral uncorrelated roughness at the 
heterojunctions. We have also calculated the conductance for a DBS in which
interface roughness is an exponentially correlated random variable, as
described previously. Figure~\ref{fig7} shows the conductance at a given
energy about the conductance resonance for the ordered case (where fluctuations
are lower), $E_c=0.1025$ eV, when the correlation 
length varies several orders of
magnitude. Physical parameters of the DBS are the same as in Fig.~\ref{fig4}.
The size of the islands is $20\,$nm and the degree of lateral disorder is
$\eta=0.3\,$nm. An average over 50 realizations of the disorder has been
performed. It can be observed that the conductance increases as the correlation
length increases. Two asymptotic regimes are clearly observed for small and
large correlation lengths, respectively. The former limit, $\xi \rightarrow 0$,
can be understood as the situation previously studied where lateral disorder is
uncorrelated. From a closed inspection of Fig.~\ref{fig7} we come to the
remarkable result that this uncorrelated limit correctly describes
transport properties whenever the correlation length do not exceed $10\,$nm.
Then we are led to an important conclusion, namely those models of electron
transport in disordered DBS based on the assumption that disorder is
uncorrelated would yield right values of the conductance even if the
correlation length is not vanishing but fairly large. There is no need to
mention that uncorrelated disordered models are much more convenient for
analytical work than correlated ones. The size of the plateau for which
correlations in the disorder do not play an essential role is governed by the
size of the islands, $\zeta$, as might be expected. This is due to the fact
that only correlations of the lateral disorder whose correlation length is
greater than the size of the islands affect electron motion. For these sizes,
the conductance increases as the correlation length becomes larger. This trend
is observed up to sizes of the order of $100\,$nm, namely the system size in
the $y$ direction, and then the second regime appears. In this regime the
limiting value of the conductance can be calculated as an average over
different ordered DBS, each one having barrier widths distributed normally.

\subsection{Compositional disorder}

Having discussed the effects of interface roughness, let us now present our
results regarding compositional disorder. To clarify the effects of local
fluctuations of the Al mole fraction, we assume that the interfaces are
perfectly flat and that the width of the different semiconductor layers are
exactly their nominal values, i.e., we neglect fluctuations of their widths. We
show typical results in Fig.~\ref{fig8}, where physical parameters are the same
as in Fig.~\ref{fig4}. The espatial extend of the region where the Al mole
fraction can be assumed locally constant was taken to be $a=10\,$nm and
$b=1\,$nm (see Fig.~\ref{fig3}). It can be observed that, for fluctuations as
large as $\Omega=0.14$, that is, height fluctuations of about $0.1\,$ eV, when
averaging over $100$ realizations the conductance is almost unchanged (only 3\%
variation). This behavior is a consequence of the small electron probability
amplitudes within the barriers, where compositional disorder is considered to
appear. In other words, even moderately high spatial inhomogeneities in the
barriers scarcely affect the electron envelope function. Thus we can conclude
that the effects of compositional disorder can be disregarded as they are much
smaller than those of lateral disorder. 

\section{Conclusions} \label{sec5}
  
In this paper we have presented a method to study electron transport in 
unintentionally disordered DBS when translational symmetry along the normal
plane to the  growth direction is broken by structural imperfections. Two
different kinds of disorder leading to electron scattering has been presented,
namely lateral disorder and compositional disorder. Lateral disorder appears as
a consequence of the roughness present at the interfaces between GaAs and
Al$_x$Ga$_{1-x}$As. Compositional disorder is due to the inhomogeneity of the
Al mole fraction occurring during the growth process of Al$_x$Ga$_{1-x}$As
epilayers. We  have shown that the main effect of the lateral disorder is to
decrease the DBS conductance. When correlations are introduced, conductance
starts improving, the larger the correlation length, the higher the
conductance. However, current shows a surprising behavior, as it is higher in
the disordered case than in the ordered one. We attributed it to an energy
shift of the transmission resonance of the DBS. Another important conclusion
regards the validity of uncorrelated disordered models; they provide right
values of the conductance whenever the actual correlation length does not
exceed the system size, thus allowing a simpler theoretical description.
Finally we have shown that the effect of compositional  disorder can be
disregarded as it is much smaller than the effect of lateral  disorder. 

\acknowledgments   

The authors want to thank V.\ A.\ Malyshev for the critical reading of the
manuscript. Work in Madrid was supported by DGI-MCyT (Project~MAT2000-0734) and
CAM (Project~07N/0075/2001). P.\ Orellana would like to thank Milenio ICM
P99-135-F and C\'atedra Presidencial de Ciencias for financial support.

\begin{figure}   
\caption{Schematic view of the sample. Regions I and III are the electrical
leads of the sample (contacts)and electrons undergo scattering processes only 
at region II.}
\label{fig1}   
\end{figure}

\begin{figure}   
\caption{Steps and islands modeling lateral disorder (roughness) at the
interface between two epilayers.}
\label{fig2}   
\end{figure}

\begin{figure}   
\caption{Schematic view of the Al fraction mesh in the barrier showing the
distribution of Al concentration at different sites.}
\label{fig3}   
\end{figure}  

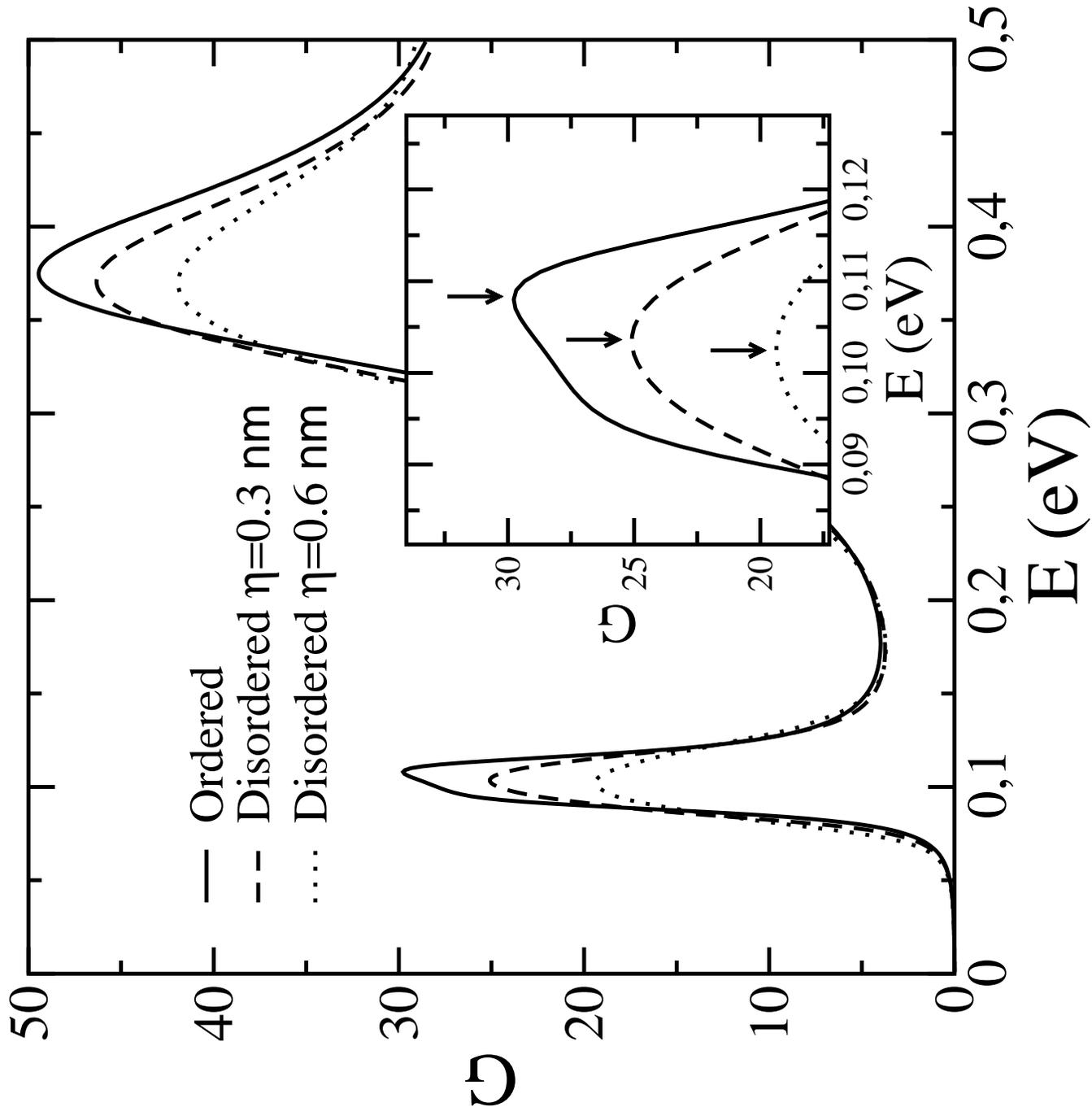
\begin{figure}   
\caption{Conductance in units of $2e^2/h$ of an ordered DBS compared with that
of a disordered DBS for two different values of the degree of lateral disorder
$\eta$. The disordered results were obtained by averaging over $100$
realizations of the disorder. The inset shows the shift of the first 
resonant peak towards lower energy on increasing the degree of disorder.}
\label{fig4}   
\end{figure}   

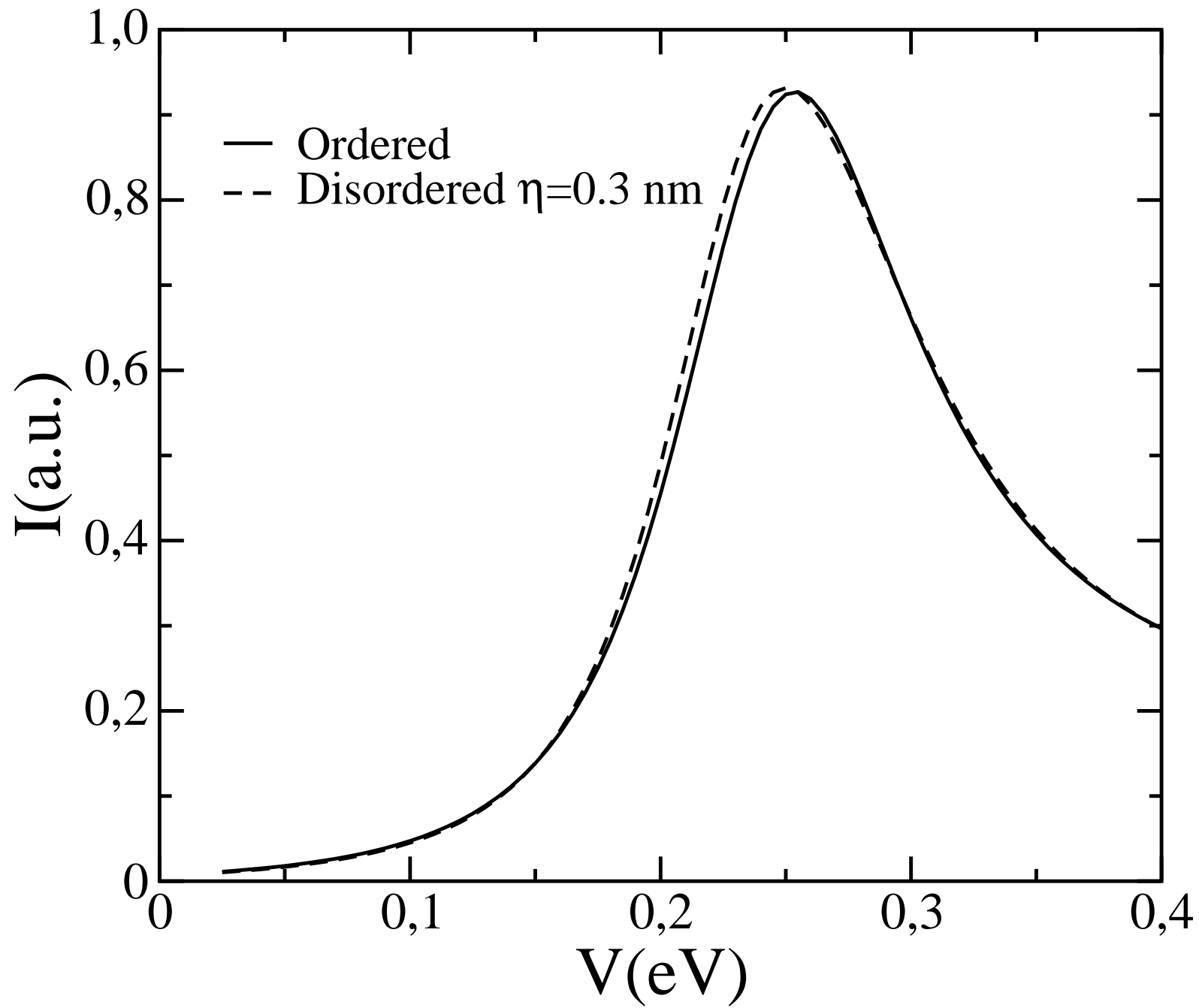
\begin{figure}   
\caption{Current in arbitrary units as a function of the applied bias along the
$z$ direction for an ordered DBS and a disordered one. Curve was calculated by
averaging over $50$ realizations of the disorder.}
\label{fig5}   
\end{figure} 

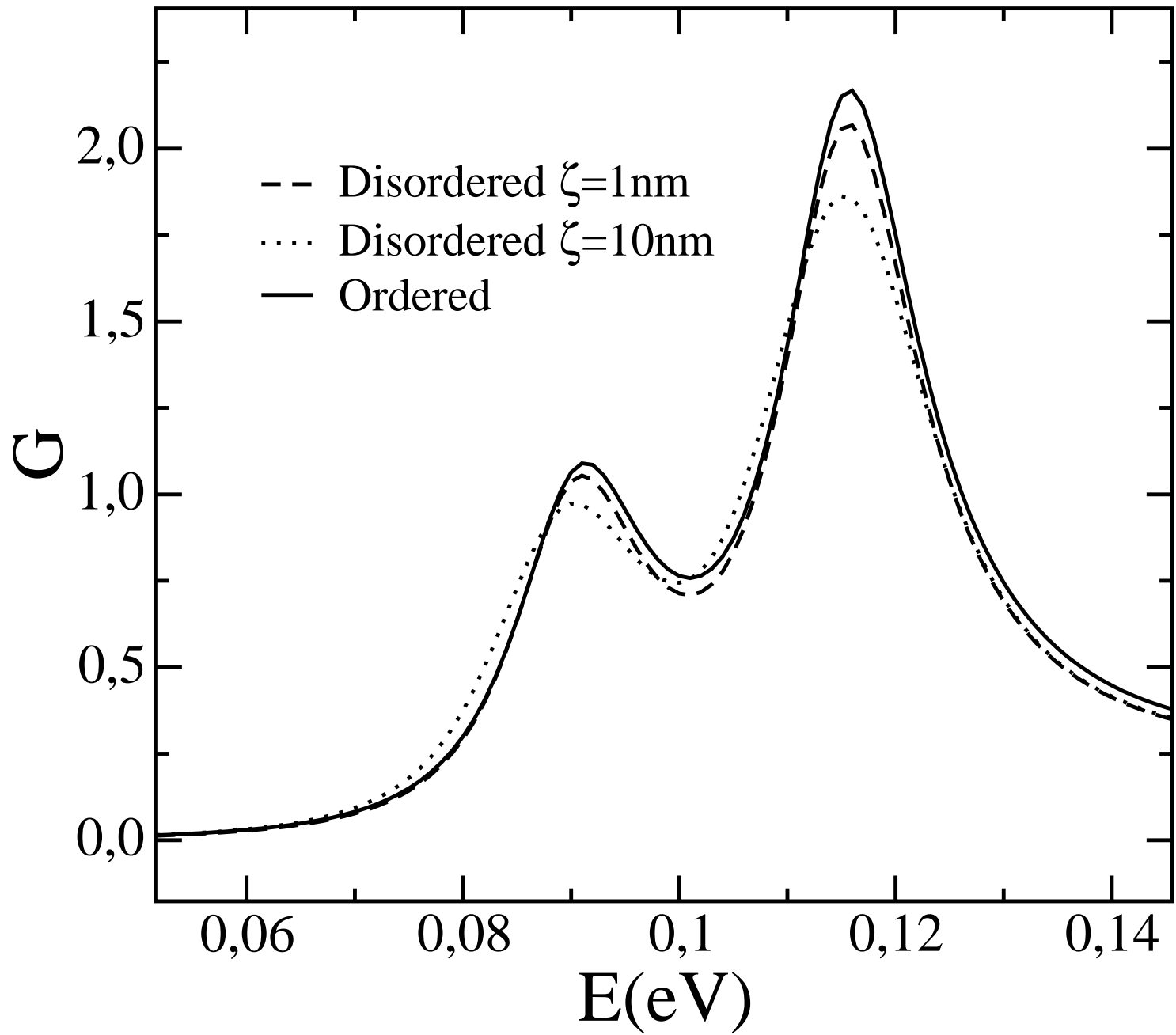
\begin{figure}   
\caption{Conductance in units of $2e^2/h$ as a function of the size of the
islands for $\zeta \sim \lambda_e$ (dotted line) and $\zeta < \lambda_e$
(dashed line). Curves were calculated by averaging over $50$ realizations of
the disorder.}
\label{fig6}   
\end{figure} 
 
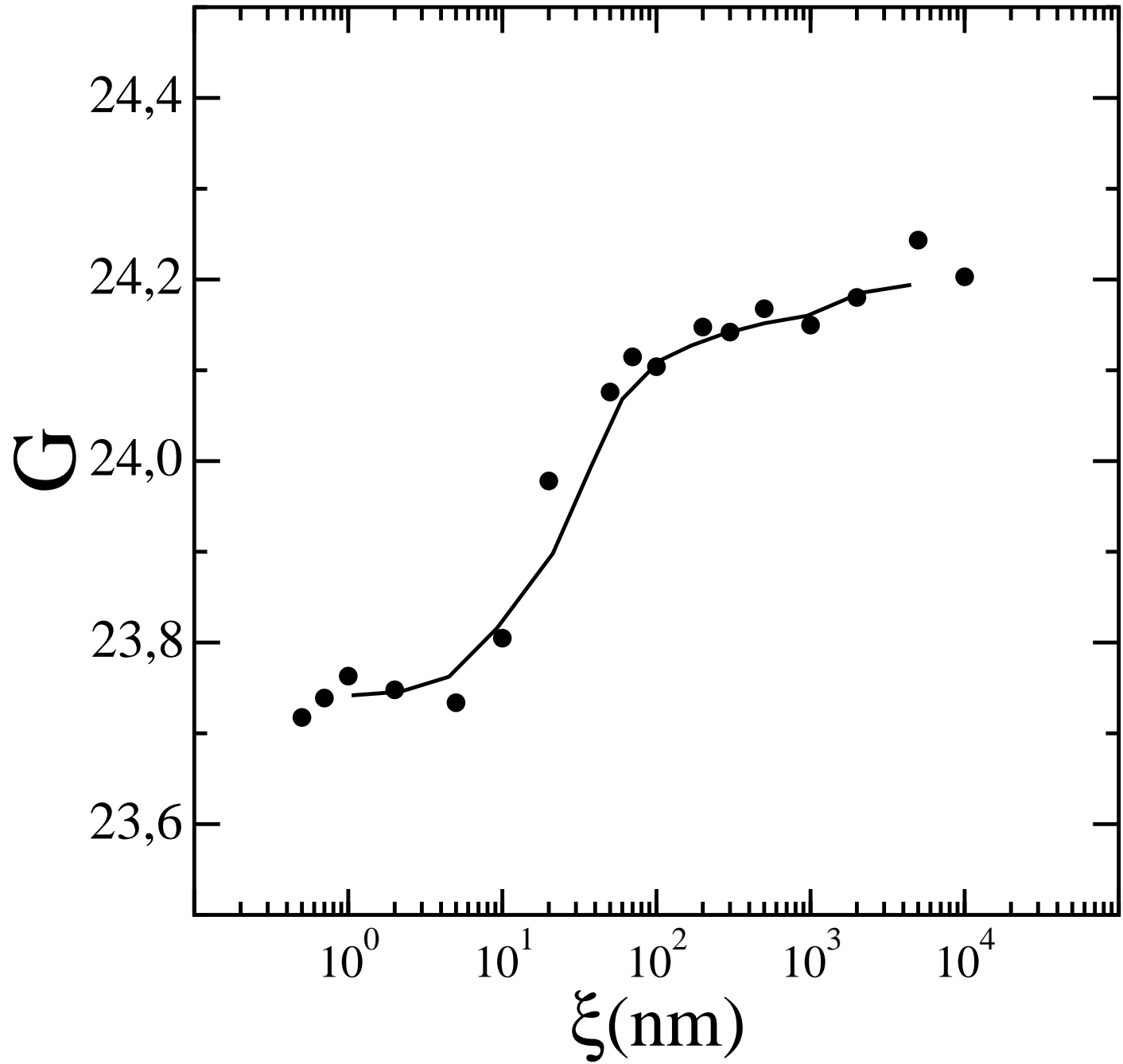
\begin{figure}   
\caption{Conductance in units of $2e^2/h$ of a disordered DBS where disorder is
exponentially correlated, as a function of the correlation length, for a fixed
energy $E_c=102.5\,$meV. Each point of the curve was calculated by averaging
over 50 realizations of the correlated disorder with fixed correlation length
$\xi$.}
\label{fig7}   
\end{figure} 

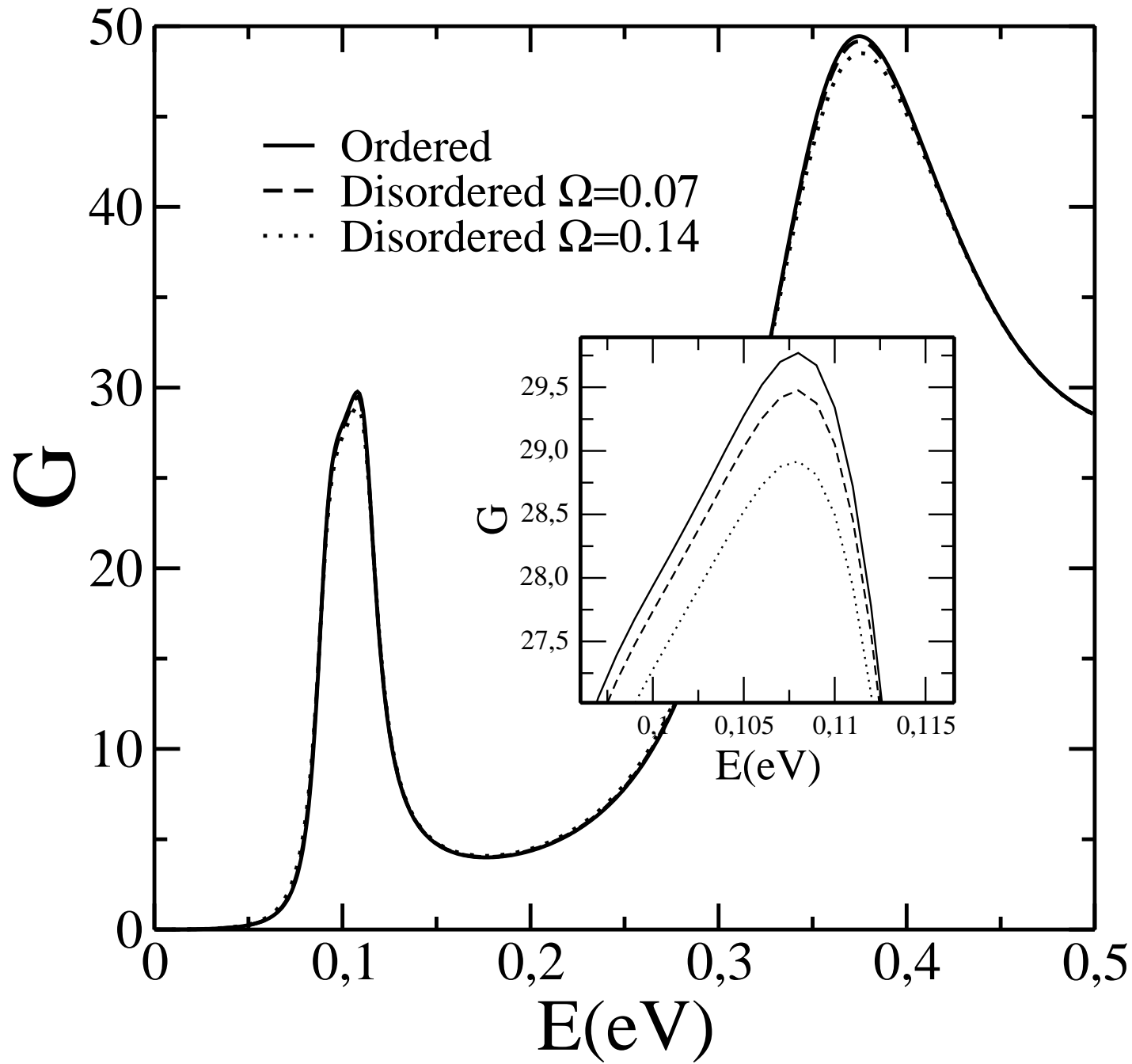
\begin{figure}   
\caption{Conductance in units of $2e^2/h$ for an ordered DBS compared with 
that of a DBS with only compositional disorder for two different values of 
$\Omega$. The inset shows enlarged view of the first resonant peak.}
\label{fig8}   
\end{figure}

\end{document}